\documentclass{article}
\usepackage{graphicx} 
\usepackage{subfig}
\usepackage{amsmath,amssymb,array}
\usepackage{booktabs}
\usepackage{float}
\usepackage{pifont}
\usepackage{framed,color,verbatim}
\definecolor{shadecolor}{rgb}{.95, .95, .95}
\newenvironment{code}%
   {\snugshade\verbatim}%
   {\endverbatim\endsnugshade}
\usepackage{ulem}
\usepackage{natbib}
\usepackage{algorithm,algpseudocode}

\usepackage{url}

\usepackage{booktabs}
\usepackage{float}
\usepackage{pifont}
\usepackage{ulem}
\usepackage{hyperref}
\hypersetup{
citecolor= blue, 
    colorlinks=true,
    linkcolor=blue, 
    filecolor=blue,      
    urlcolor=blue,
    pdftitle={Overleaf Example},
    pdfpagemode=FullScreen,
    }

\title{Testing for a general changepoint in psychometric studies:
changes detection and sample size planning}
\author{Nicoletta D'Angelo}
\date{Department of Economics, Business, and Statistics University of Palermo\\
Palermo, Italy}

\begin{document}

\maketitle

\begin{abstract}
This paper introduces a new method for change detection in psychometric studies based on the recently introduced pseudo Score statistic, for which the sampling distribution under the alternative hypothesis has been determined.
Our approach has the advantage of simplicity in its computation, eliminating the need for resampling or simulations to obtain critical values. Additionally, it comes with a known null/alternative distribution, facilitating easy calculations for power levels and sample size planning.
The paper indeed also discusses the topic of power analysis in segmented regression, namely the estimation of sample size or power level when the study data being collected focuses on a covariate expected to affect the mean response via a piecewise relationship with an unknown breakpoint.
We run simulation results showing that our method outperforms other Tests for a Change Point (TFCP) with both normally distributed and binary data and carry out a real SAT Critical reading data analysis. 
The proposed test contributes to the framework of psychometric research, and it is available on the Comprehensive R Archive Network (CRAN) and in a more user-friendly Shiny App, both illustrated at the end of the paper.
\end{abstract}%

\section{Introduction}


In recent years, item-level response time data has become more accessible due to computer-based testing and online survey data collection methods \citep{cheng2021application}. This has led to a significant rise in interest within the field of psychometric research \citep{lee2014using}.

In parallel, Tests for a Change Point (TFCP) have emerged for their potential uses in psychometrics \citep{chen2012parametric,hawkins2003changepoint}. Originating from the field of statistical quality control \citep{allalouf2007quality,montgomery2020introduction,von2012use}, TFCPs are methods intended to detect whether there has been any change in the parameters underlying a sequence of random variables. Specifically, TFCPs focus on finding the moment in time when the statistical model or its parameters underlying a sequence of observations have changed in some fashion \citep{montgomery2020introduction}. These methods involve testing the null hypothesis of no change against the alternative hypothesis that a change has occurred after a certain observation.

TFCPs have been successfully applied to detect such unusual change psychometric-related problems. Some are worth mentioning: \cite{lee2013monitoring} used a TFCP to detect an unusual change in the mean score of a sequence of administrations of an international language assessment, and  \cite{shao2015change}, which used a TFCP to detect speededness in non-adaptive tests. 

Moreover, \cite{sinharay16} demonstrated how a TFCP can be used to detect an abrupt change in the test performance of examinees during a Computerized adaptive test (CAT). By comparing the performances of the new statistics with those of four existing TFCPs, the author shows that the TFCPs appeared promising for assessment of person fit for CATs.

More recently, \cite{cheng2021application} proposed the usage of two test statistics based on changepoint analysis, namely the likelihood ratio test and Wald test employed by \cite{lee2013monitoring} and \cite{shao2015change} respectively, to detect test speededness.
Indeed, one notable application of response time has been the detection of aberrant response
behavior \citep{van2013speededness}. 
Conversely, to other common approaches to detect test speededness using response time data, \citeauthor{cheng2021application}'s proposal does not examine how an individual response time pattern deviates from the group behaviour or model-implied behaviour, but it concerns itself with intraindividual change during the test-taking process.
Finally, \cite{zhu2023bayesian} recently proposed a changepoint analysis procedure using response times to detect abrupt changes in examinee speed, which may be related to aberrant responding behaviours in the Bayesian context.

The differences in the type of data and test statistics employed leave no doubt about the great set of psychometric problems which TFCPs can solve. 

All the above-mentioned approaches, however, lack a known null and alternative distribution, introducing, therefore, uncertainty, subjectivity, and challenges in both the interpretation of results and the design of robust, reproducible experiments.

Furthermore, conducting power analysis, essential for determining the ability of a test to detect true effects, becomes less reliable without a known distribution.


Borrowing terminology from the basics of Statistical Inference theory, the \textit{power} of a study is the probability of detecting a significant covariate effect on the response. Roughly speaking, the power depends mostly on the sample size, the statistic test being used with fixed type-I error probability, and some settings related to the specific problem and model under investigation.

For instance, the researcher has to specify the effect size via Cohen’s $d$ when the study involves the traditional mean comparisons between two groups or the expected correlation coefficients when the study focuses on the association between two variables \citep{cohen2013statistical}. 

From a practical point of view, the most important goal of \textit{power analysis} is to estimate the sample size when a desired level of power is fixed. Namely, the investigator anticipates a certain effect size, sets a significance level $\alpha$, and then specifies the amount of power they desire. Then, power analysis is used to determine the sample size $n$, which is necessary to meet their specifications. Having some, even rough, idea of power/sample size is crucial for better planning the study and efficient resource allocation, namely avoiding waste of time and costs. Experimental results with too low statistical power will lead to invalid conclusions about the meaning of the results, and therefore a minimum level of statistical power is commonly sought.

In addition to the traditional and well-known mean comparisons or correlation, power analysis has been studied in different scenarios, including survival analysis in epidemiological studies \citep{powerSurvEpi}, random effects and mixed models \citep{green2016simr} and general power analysis methods in genetic studies \citep{powerGWASinteraction}. Moreover, \cite{Moerbeek2022} investigates the power needed to detect differential growth for linear–linear piecewise growth models. 

However, to the best of our knowledge, there is no paper dealing with power analysis when the main relationship under investigation is segmented, namely a continuous covariate affecting the mean response via two straight lines connected at an unknown covariate value, the so-called breakpoint or changepoint, where the effect changes abruptly.

In psychological/psychometric research, changepoint models and relevant applications have been discussed by \citet{sinharay16, sinharay17} and more recently by \citet{cheng2021application} to detect test speededness using response time data. 


Following up on such works, we propose an alternative to the most known TFCPs, based on the pseudo Score statistics proposed by \cite{muggeo2016testing}. The aim of this research work is to propose a test which is simple to compute, without resampling or simulations requested to get critical values, and with a known null/alternative distribution which makes the computation of the power level and/or sample size planning straightforward. 


The structure of the paper is as follows.
First, in Section \ref{sec:tests}, we review the theory for TFCP analysis.
Secondly, we introduce the pseudo Score statistic and illustrate how it can be used to test the existence of a breakpoint in segmented regression models in Section \ref{sec:seg}.
In Section \ref{sec:sims}, we then assess the performance of our proposal through a simulation study in terms of rejection rates, comparing it to the performance of two TFCPs: the likelihood ratio and Wald tests.
Furthermore, we carry out an analysis of real SAT Critical reading data in Section \ref{sec:analysis}, to show the applicability of the method.
After having assessed its performance, Section \ref{sec:pow.seg} illustrates how to employ the proposed test statistics to conduct power analysis.
Finally, we present the new \texttt{R} function available on the \texttt{segmented} package \citep{muggeo2008segmented} of the Comprehensive R Archive Network (CRAN), and a Shiny App (\url{https://uy1z3u-nicoletta0d0angelo.shinyapps.io/power_seg/}), which implements the proposed methodology. Indeed, all the analyses are available from the author and carried out through the statistical software \texttt{R} \citep{R}. The paper ends with some conclusions.

\section{Tests for a Change  Point}
\label{sec:tests}

In this section, we review the Tests for a Change  Point, henceforth called TFCP.
We shall suppose that the data consists of the observations $Y_1, Y_2, \dots, Y_n$, obtained at a sequence of time points $i = 1, 2, \dots, n$. We further assume that the $Y_i$'s come from an underlying statistical model, depending on some parameters. A TFCP is commonly employed to determine whether there exists a time point $\psi$ such that the model parameter underlying the sub-sequence $Y_1, Y_2, \dots, Y_{\psi - 1}$ is statistically different from the one underlying the sub-sequence $Y_\psi, Y_{\psi + 1}, \dots, Y_{n}$. The time point $\psi$ is referred to as \textit{changepoint}, or \textit{breakpoint}.

Among the several formulations of TFCPs, we follow the most relevant to psychometric problems discussed in \cite{andrews1993tests}, as follows.

Formally, let $Y_1, Y_2, \dots, Y_n$ be independent random variables, with probability function $f_i(Y_i; \tau_1, \eta)$ for $i = 1, 2, \dots, \psi - 1$, and 
$f_i(Y_i; \tau_2, \eta)$ for $i = \psi, \psi + 1, \dots, n$. The parameters $\tau_1$ and $\tau_2$ are those of interest, while $\eta$ is the nuisance parameter. The reason why the $X_i$s have not been assumed identically distributed is that they could, for instance, denote the scores on items of the same examinee.

A TFCP typically tests either 
\begin{equation}
	\begin{cases}
		\mathcal{H}_{0} : & \tau_1=\tau_2\\
		\mathcal{H}_{1} : & \tau_1\neq \tau_2
	\end{cases}
	\label{eq:syst1}
\end{equation}
or the alternative one-sided hypothesis $\mathcal{H}_{1}: \tau_1 > \tau_2$ or  $\mathcal{H}_{1}: \tau_1 < \tau_2$.

We address the most relevant problem in psychometrics, which is that of unknown $\tau_1, \tau_2, \eta,     \text{ and }  \psi$.

Next, we review the appropriate test statistics depending on the distribution of the $f_i$s, following the general overview of \cite{sinharay17}.

\subsection{Normally distributed observations}

The following method tests whether the means in two parts of a record are different (for an unknown
changepoint). The test assumes that the data are normally distributed, with $\tau_1$ and $\tau_2$ as the means and $\eta$ as the common variance, which is a common setting in TFCPs.

Therefore, the generalized Likelihood Ratio Test (LRT) of the hypothesis system \eqref{eq:syst1} can be performed using the following test statistics:
\begin{equation}
     T_{\text{max},n} = \max_{1 \leq j \leq n - 1}{|t_{jn}|}
     \label{eq:t_max}
\end{equation}
where 
\begin{equation*}
    t_{jn} = \sqrt{\frac{j(n-j)}{n}}\frac{\Bar{Y}_{jn} - \Bar{Y}^*_{jn}}{s_{jn}},
    \end{equation*}
    \begin{equation*}
    \Bar{Y}_{jn} =\frac{1}{j}\sum_{i=1}^{j}Y_i, \quad
    \Bar{Y}^*_{jn} =\frac{1}{n-j}\sum_{i=j+1}^{n}Y_i, 
    \end{equation*}
    \begin{equation*}
\text{and}      \quad 
s^2_{jn}= \frac{\sum_{i=1}^{j}(Y_i - \Bar{Y}_{jn})^2 + \sum_{i=j+1}^{n}(Y_i - \Bar{Y}^*_{jn})^2}{n-2}.
\end{equation*}
The null hypothesis is rejected if $T_{\text{max},n}$ is larger than an appropriately chosen critical value $h_n$.

\subsection{Non-normal distributed observations}

If the $Y_i$s are assumed to follow a binary distribution, denoting, for instance, scores on binary items, \cite{andrews1993tests} and \cite{csorgHo1997integral} show that one can use the following LRT statistics
\begin{equation}
    L_{\text{max},n} = \max_{n1 \leq j \leq n - n1}{L_{jn}}
    \label{eq:LRT}
\end{equation}
where
\begin{equation*}
\begin{split}
    L_{jn} &= 2\{ L_{j1}(\hat{\tau}_{1j}, \hat{\eta}_a; Y_i, i = 1, 2, \dots, j) + L_{j2}(\hat{\tau}_{2j}, \hat{\eta}_a; Y_i, i = j + 1, j + 2, \dots, n)   \\
    &- L(\hat{\tau}_{0}, \hat{\eta}_0; Y_i, i = 1, 2, \dots, n)\},
\end{split}
\end{equation*}
and for example the log-likelihood of $Y_1, Y_2, \dots, Y_j$ at $\tau_{1j}$ is
\begin{equation*}
\begin{split}
    L_{j1}(\tau_1, \eta; Y_i, i = 1, 2, \dots, j)&= \sum_{i=1}^{j}\log{f_i(Y_i; \tau_1, \eta)} \\ 
    &= \sum_{i=1}^{j}[Y_i\log{P_i(\tau_{1j})} + (1-Y_i)\log{1-P_i(\tau_{1j})}].
    \end{split}
\end{equation*}

The statistics \eqref{eq:LRT} is employed to test the null hypothesis in \eqref{eq:syst1} for all $j$s versus the alternative of a change between items $n_1$ and $(n -n_1)$. Indeed, to increase the stability of the test, \cite{andrews1993tests} recommends setting $n_1$ to the nearest integer to $0.15n$, which restricts the changepoints to roughly the middlemost $70\%$ of the observations. 

Also, in this case, one rejects the null hypothesis if $L_{\text{max},n}$ is larger than an appropriately chosen critical value $h_n$.

\section{The pseudo Score statistic 
}
\label{sec:seg}

We now introduce the terminology about segmented regression models and review the pseudo Score statistics proposed in this paper as a tool for performing changepoint detection, alternatively to the previously reviewed TFCP, and for conducting power analysis.

The segmented regression model with a single 
changepoint $\psi$ in the covariate $z$ is 
\begin{equation}
g(\mathbb{E}\left[Y|x_i,z_i \right])=x_i^T\boldsymbol{\beta}+ \delta \varphi \left(z_i,\psi\right) \qquad i = 1, 2, \ldots, n,
\label{eq:seg}
\end{equation}
where $\mathbb{E}[\cdot]$ denotes the expected value, 
$Y$ is the response variable, $x_i^T$ is the possible vector of additional non-segmented covariates related linearly to the mean response with associated parameter vector $\boldsymbol{\beta}$, and $z$ is the segmented variable with a piece-wise linear relationship: i.e., the $z$-effect changes by $\delta$ at the unknown $\psi$ in the covariate range \citep{muggeo2003estimating}.  The expected value of the response is linked to the right-hand side of equation  \eqref{eq:seg}, i.e. the \textit{linear predictor},  through a \textit{link function} $g(\cdot)$, a monotone function which ensures admissible values of the predicted responses.

Note that this formulation allows, in addition to the inclusion of potential other non-segmented covariates, for the segmented variable $z$ not to necessarily be the sequence of time points observed for the $Y_i$s. 

In general, $z_i$ could be any other variable external to the phenomenon under analysis. 
For instance, \cite{priulla2021analysis} considered the number of university credits earned during the first year as a good predictor of the regularity of the career and, therefore, considered it as a potential segmented covariate, whose effects on the probability of getting the bachelor's degree within 4 years could change.
Other examples of application fields include epidemiology, occupational medicine, toxicology, and ecology \citep{ulm1991statistical,betts2007thresholds}.

In the context of changepoint detection, there are several cases covered by the function $\varphi$ and, in turn, by
model \eqref{eq:seg}.

To parallel the cases considered by the TFCPs, we consider in this paper the segmented regression model with a single \textit{discontinuous} 
changepoint $\psi$ in the covariate $z$, also called \textit{jump-points} model, an no additional non-segmented variables, as follows:
\begin{equation}\label{eq:regress}
g(\mathbb{E}\left[Y|z_i \right])=\beta+ \delta I \left(z_i>\psi\right),
\end{equation}
and include the only segmented covariate as the sequence of the time points observed for the $Y_i$s, that is, set $z_i = i$. The link function $g(\cdot)$ will change depending on the distribution assumed for the $f_i$s.

Testing for the existence of a changepoint means to test for the following system of hypothesis $\mathcal{H}_{0}:  \delta=0$ vs. $\mathcal{H}_{1}: \delta\neq0$, where the 
alternative $\mathcal{H}_{1}$ can also be unidirectional.


Using the setting of probabilistic coherence of de Finetti, \citet{muggeo2016testing} proposes to use the following pseudo Score statistic
\begin{equation} \label{eq:score}
    \bar{\varphi}^{T}(I_{n}-A)y
\end{equation}
where $I_{n}$ the identity matrix, $y$ the observed response vector, and $A$  is the hat matrix under the null hypothesis of no changepoint. $\bar{\varphi}=(\bar{\varphi}_{1},...,\bar{\varphi}_{n})^{T}$ is the vector of the segmented term $\varphi(z_i,\psi)$ averaged over the range of $Z$ i.e. $\bar{\varphi}_i=K^{-1}\sum_{k=1}^{K}\varphi(z_{i},\psi_{k})$ using $K$ fixed values $\{\psi_k\}_{k=1,...,K}$. 

The expected value of \eqref{eq:score} is zero under $\mathcal{H}_{0}$, and its variance can be easily obtained, therefore the test statistic based on the pseudo Score   \eqref{eq:score} is 
\begin{equation}\label{eq:Test}
s_0=\frac{\bar{\varphi}^{T}(I_{n}-A)y}{\{\sigma^2 \bar{\varphi}^{T}(I_{n}-A)\bar{\varphi}   \}^{1/2}} \stackrel{\mathcal{H}_0}{\sim} \mathcal{N}(0,1)
\end{equation}
When the unknown variance $\sigma^2$ is replaced by its estimate under the null or the alternative hypothesis or the response variable does not follow a Normal distribution, the null distribution in \eqref{eq:Test} is asymptotically a standard Normal, but convergence is very fast, and results are accurate even in modest sample sizes.

\section{Simulation study}
\label{sec:sims}

This section is devoted to assessing the performance of the pseudo Score test \eqref{eq:Test} in terms of rejection rates and its comparison with benchmark TFCPs.

\subsection{Normal data}

We simulated from twelve different scenarios, generating normally distributed data with three different true values of the mean difference, namely $\delta=\{.25,.5,1\}$, and considering four different sample sizes $n=\{20,30,40,50\}$. We include only one segmented covariate, taking equispaced values ranging from $0$ to $1$. 
The jump-points model \eqref{eq:regress} used for the simulations is
$$y_i=\beta+\delta I(i > \psi)+\epsilon_i,$$  
 considering the intercept $\beta=2$, the true changepoint $\psi = 0.5$, identity link function $g(\cdot)$, and i.i.d. standard Gaussian errors with $\sigma = 0.3$ standard deviation.
As for the hypothesis testing, we fix $\alpha=0.05$.

As we are dealing with normally distributed data, we compare our proposal with the test in \eqref{eq:t_max}.
Its critical values are provided in \cite{worsley1979likelihood} for the equivalent test statistics
$W_{\text{max},n} = \frac{(n-2)^{0.5}T_{\text{max},n}}{(1-T_{\text{max},n}^2)^{0.5}}$,
and are reported in Table \ref{tab:W}. 

\begin{table}[htb!]
	\centering
	\caption{Table of critical values of W \citep{worsley1979likelihood}}
	\begin{tabular}{cccc}
	\toprule
&\multicolumn{3}{c}{$\alpha$}\\
n&0.10&0.05&0.01\\
	\midrule
	10 &3.14 & 3.66 & 4.93\\
	15 &2.97 & 3.36 & 4.32\\
	20 &2.90 & 3.28 & 4.13\\
	25 &2.89 & 3.23 & 3.94\\
	30 &2.86 & 3.19 & 3.86\\
	35 &2.88 & 3.21 & 3.87\\
	40 &2.88 & 3.17 & 3.77\\
	45 &2.86 & 3.18 & 3.79\\
	50 &2.87 & 3.16 & 3.79\\
	\bottomrule
	\end{tabular}
\label{tab:W}
\end{table}

\subsection{Binary data}

For non-normal distributed data, we follow the simulation setup in \cite{sinharay16}, assuming a Computerized adaptive test (CAT) with $n$ dichotomous items being administered to an examinee whose true ability is denoted by $\theta$. Let $Y_i, \quad i = 1, \dots, n$ be the dichotomous examinee's score obtained on the $i$-th item.
The probability of a correct answer on item $i$ is denoted by $P_i(\theta)$. 
The simulated data are obtained from a three-parameter logistic model
\begin{equation}
    P_i(\theta) = c_i + (1 - c_i) \frac{\exp{a_i(\theta - b_i)}}{1 + \exp{a_i(\theta - b_i)}},
    \label{eq:rasch}
\end{equation}
where $a_i, b_i, \text{ and } c_i$ represent the slope, difficulty, and guessing parameters on item $i$, respectively.
Indeed, the data are simulated from a Rasch model \citep{rasch1993probabilistic}, given by equation \eqref{eq:rasch} with $a_i=1$ and $c_i = 0$, making it basically a logit model, that is, equation \eqref{eq:regress} with logit link function.

To compute the Type I error rates (that is, under the hypothesis of no changepoint), the true difficulty parameters $b_i$ are drawn from a standard Normal distribution.

To compute the power (that is, under the hypothesis of the existence of a changepoint), the true abilities for the two halves of the CAT (denoted by $\theta_1$ and $\theta_2$) are obtained as follows: for each examinee, $\theta_1$ is simulated from a standard normal distribution, and $\theta_2$ is set as $\theta_1 + \delta$. In particular, we consider $\delta = \{1, 2, 3\}$, where of course, positive values of $\delta$ indicate an improvement in the performance in the second half of the test. 

For any test length, the rejection rate is computed on 1000 model-fitting score patterns.
In particular, we consider four levels of test length $n$, involving 20, 30, 40, and 50 items. 
The true changepoints are 11, 15, 21, and 25 for the four tests, respectively, similar to those in \cite{sinharay17}.

For this scenario, we compare our results with the test in \eqref{eq:LRT}.
The critical value for $L_{\text{max},n}$ is 8.85 for $n1/n = 0.15$ and $\alpha = 0.05$ \citep{andrews1993tests}.

\subsection{Results}

Tables \ref{tab:powern} and \ref{tab:power}  report the rejection rates obtained over 1000 simulations each for the normal and binary data, respectively. 

We denote by P.Score the columns with the rejections rates obtained from the application of our proposal \eqref{eq:Test} and by W and L those following the tests \eqref{eq:t_max} and \eqref{eq:LRT}, respectively.

Under the null hypothesis, that is, in the absence of a change ($\delta = 0$), P.Score shows similar rejection rates compared to W. However, for binary data, it stays close to a 0.05 value, contrary to L, which appears to depend on the sample size $n$.

Moving to the alternative hypothesis, we find no differences in the rejection rates of P.Score and W when both sample size and the mean difference $\delta$ are small. This behaviour is observed for this case uniquely, while in general, P.Score always achieve slightly better rejection rates compared to both W and L. 

Overall, the performance of P.Score aligns with those of W and L. This means that reasonable rejection rates under the null hypothesis are observed, close to the chosen significance level of 0.05.  It also means that, as expected, higher rejection values under the alternative hypothesis increase with the sample size and the magnitude of the mean difference of the change.

\begin{table}[htb!]
	\centering
	\caption{Rejection rates of the tests at a $.05$ significance level over 1000 simulations for normal data}
	\begin{tabular}{c|cc|cc|cc|cc}
	\toprule
&\multicolumn{2}{c}{$\delta=0$}&\multicolumn{2}{c}{$\delta=.25$}&\multicolumn{2}{c}{$\delta=.5$}&\multicolumn{2}{c}{$\delta=1$}\\
$n$&P.Score&W&P.Score&W&P.Score&W&P.Score&W\\
	\midrule
	20&  0.044 & 0.048  &  0.081 & 0.081 & 0.126 & 0.101 & 0.429 & 0.336 \\
	30&   0.042 & 0.057 &  0.085 & 0.080 & 0.202 & 0.162 & 0.584 & 0.544  \\
	40 &   0.052 & 0.044 &  0.099 & 0.074 & 0.262 & 0.220 & 0.744 & 0.698 \\
	50& 0.054 & 0.052 &0.112 & 0.088 & 0.314 & 0.244 & 0.813 & 0.765 \\
	\bottomrule
	\end{tabular}
\label{tab:powern}
\end{table}

\begin{table}[htb!]
	\centering
	\caption{Rejection rates of the tests at a $.05$ significance level over 1000 simulations for binary data}
	\begin{tabular}{ccc|cc|cc|cc}
	\toprule
&\multicolumn{2}{c}{$\delta=0$}&\multicolumn{2}{c}{$\delta=1$}&\multicolumn{2}{c}{$\delta=2$}&\multicolumn{2}{c}{$\delta=3$}\\
$n$&P.Score&L&P.Score&L&P.Score&L&P.Score&L\\
	\midrule
	20& 0.054  & 0.031  &0.130  &  0.087 &0.294  & 0.222  & 0.433 &  0.385\\
	30& 0.050  & 0.042 &0.168  & 0.141 & 0.421  & 0.378 & 0.654  & 0.624 \\
	40& 0.045  & 0.058&0.201  & 0.175 &0.530  & 0.475  &0.807  & 0.783 \\
	50& 0.050  & 0.064  &0.222  & 0.189  & 0.618  & 0.587 &0.872  & 0.866 \\
	\bottomrule
	\end{tabular}
\label{tab:power}
\end{table}

Note that similar analyses have been run with negative values of $\delta$, showing no difference in the results and conclusions, and therefore are not reported here for brevity.

\section{Real data analysis}
\label{sec:analysis}

In this section, we analyse the same data in \cite{sinharay17}, with the aim of estimating a changepoint in the mean scores on SAT Critical reading.
We consider the 2015 total-group profile report for College-bound seniors published
by the College Boar, and available on the website
\url{https://secure-media.collegeboard.org/digitalServices/pdf/sat/total-group-2015.pdf}.

\begin{figure}[!htb]
    \centering
	\subfloat{\includegraphics[width=.5\textwidth]{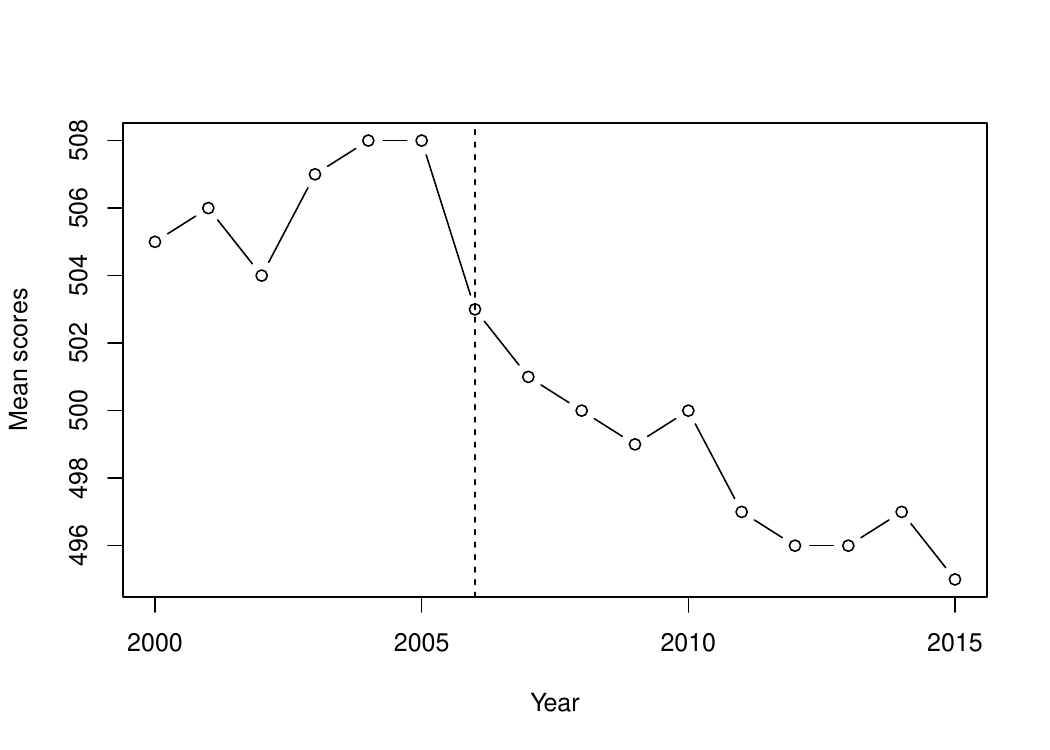}}  
	\subfloat{\includegraphics[width=.5\textwidth]{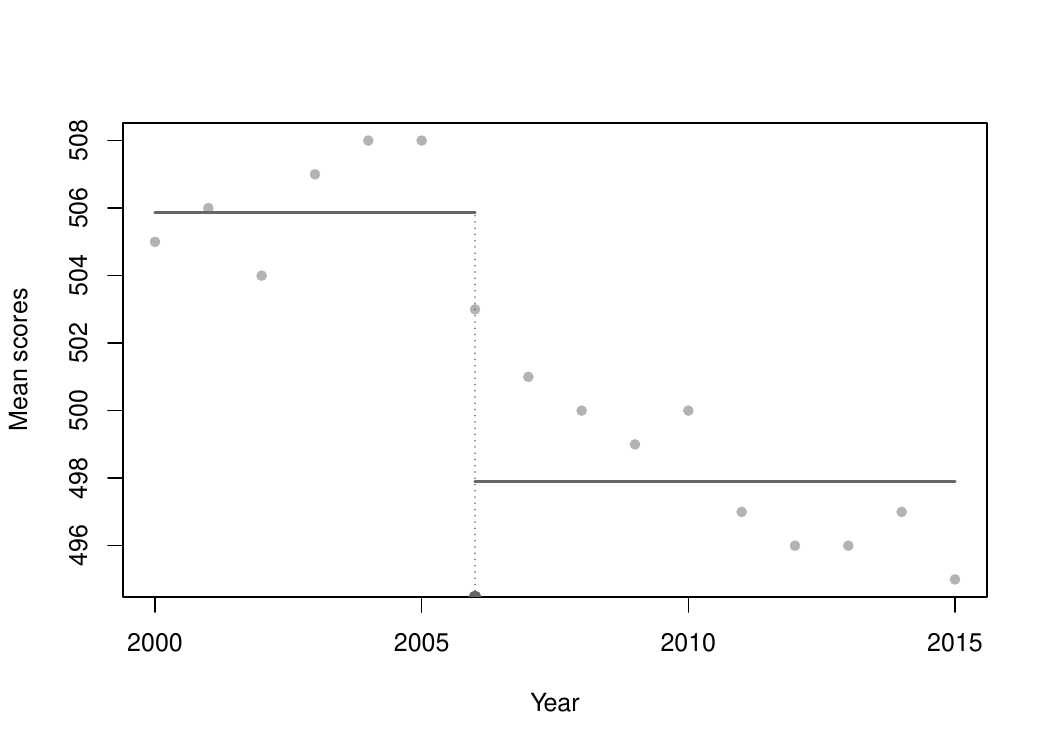}}
    \caption{Mean scores on SAT critical reading for the total group between 2000 and 2014. \textit{Left panel}: W; \textit{Right panel}: P.Score.}
    \label{fig:appl1}
\end{figure}

In particular, we analyse the mean scores on SAT Critical reading
for the total group between the years 2000 and 2015, which are shown in Figure \ref{fig:appl1}.

We are interested in testing the null hypothesis of no change in the means of the $Y_i$'s against a two-sided alternative hypothesis. Since they assumed the $Y_i$’s to be independent and to follow a normal distribution with unknown means and variance, they used the test statistic $T_{\text{max},n}$. The critical values at significance level $\alpha = 0.05$ are $3.34$ for $n = 16$.
The value of $T_{\text{max},n}$ for these data was found to be equal to $7.65$, which is much larger than the critical value $3.34$ at level $\alpha = 0.05$.
Thus, the value of $T_{\text{max},n}$ is statistically significant at level $\alpha = 0.05$—so it can be concluded that a statistically significant change occurred in the mean. The estimated changepoint is 2006 for both the W and the P.Score statistics. Particularly, the P.Score method yields a p-value of 0.0005. 

Therefore, we can conclude that in real data applications, our method's results are coherent with the standard TFCP procedures.
In particular,  several sources, such as CBS News, reported a sharp
drop in the SAT scores in 2006, making the estimated changepoint reasonable to interpret and put into context.

\section{Power analysis 
}
\label{sec:pow.seg}

In this section, we present the applicability of the proposed test statistics to conduct power analysis, having already assessed its performance under both the null and alternative hypothesis of the presence of a changepoint.

Assume now a segmented regression model with a single \textit{non-discontinuous} changepoint $\psi$ in the segmented covariate $z$, as follows:
\begin{equation*}
g(\mathbb{E}\left[Y|x_i,z_i \right])=x_i^T\beta+\delta\left(z_i-\psi\right)I\left(z_i>\psi\right).
\end{equation*}

Basically, we set $\varphi(z_i,\psi)$ of model \eqref{eq:seg} to $\delta\left(z_i-\psi\right)I\left(z_i>\psi\right)$, denoted by $\delta(z-\psi)_+$ henceforth.

In order to conduct power analysis, we need to derive the sampling distribution of the pseudo Score test statistic   \eqref{eq:score} under $\mathcal{H}_1$. 

The scale (variance) and shape (Normality) are preserved under $\mathcal{H}_1$, but when the changepoint $\psi$ does exist, $y=X\beta+\delta(z-\psi)_+$ and therefore
\begin{eqnarray} \label{eq:Ex1}
\mathbb{E}_1[\bar{\varphi}^{T}(I_{n}-A)y] &=& \bar{\varphi}^{T}(I-A)X\beta+ \delta \bar{\varphi}^{T}(I_n-A) (x-\psi)_+\\ \nonumber
 &=& \delta \bar{\varphi}^{T}(I_n-A) (z-\psi)_+
\end{eqnarray}



Given the null and alternative sampling distributions of the pseudo Score statistic $s_0$, it appears straightforward to carry out power analysis. 

As usual, given the type I error probability $\alpha$, the slope difference $\delta$, and the changepoint $\psi$, computations just require the Normal distribution function $\Phi(\cdot)$. 

The formula \eqref{eq:Ex1} reports the expected value under $\mathcal{H}_1$: the larger the expected value, the farther the alternative distribution, and the higher the power. Looking at $\mathbb{E}_1[s_0]$, it appears clear that, as expected, higher power is obtained as $\delta$ increases (in absolute value) since the segmented relationship gets more clear-cut. However, $\mathbb{E}_1[s_0]$, and the power, in turn, also depends on $\psi$, the segmented covariate distribution, and the design matrix $X$, namely possible additional covariates understood to affect the mean response. 


In general, $\mathbb{E}_1[s_0]$ depends on the segmented covariate distribution. It decreases as $\psi$ moves to the boundaries of the segmented covariate, and more additional covariates are accounted for. These factors, along with $n$, $\delta$ and $\sigma$, affect the power of the segmented regression model.


\subsection{Software implementation}

We devote this section to illustrating the implemented \texttt{R} code to perform power analysis based on the pseudo Score statistics.

Power analysis based on the Score statistic in segmented regression can be carried out via the function \texttt{pwr.seg()} which is included in the \texttt{segmented} package \citep{muggeo2008segmented}. 

\paragraph{Power computation}

For some settings specified ($\psi=0.6$, $\delta=0.5$, $z_i=i/n$, $\sigma=.1$, and $n=100$), the power (assuming the default values \texttt{alpha=0.01} and \texttt{alternative="two.sided"}) is easily obtained by typing 
\begin{code}
> pwr.seg(n = 100, z = "1:n/n", psi = .6, d = .5, s = .1)
Est. power: 0.749 
\end{code}
The segmented covariate is specified in the argument \verb|z| via a string indicating the known quantile function having \verb|'p'|, \verb|'n'| and its parameter values as arguments. In addition to \verb|"1:n/n"| (default), some examples are \texttt{"qnorm(p, 2, 1.5)"}, \texttt{"qexp(p,2.5)"}, or \texttt{"qbeta(p,1,2)"}. The \verb|psi| value has to be within the covariate range. Yet another option is to pass a numerical vector representing the actually segmented covariate.

It is probably instructive to compare the above-returned power of $0.749$ with the actual power based on $1000$ Monte Carlo replicates. Namely, for each simulated sample, we apply the Score test \eqref{eq:score} via the function \verb|pscore.test()| in the package \verb|segmented|,

\begin{code}
> set.seed(123)
> n <- 100
> z <- 1:n/n
> p <- rep(NA, 1000)
>
> library(segmented)
> for(i in 1:1000){
+  y<- .5 * pmax(z - .6,0) + rnorm(n) * .1
+  o <- lm(y ~ z)
+  p[i] <- pscore.test(o, dispersion = s^2)$p.value
+ }
\end{code}

We then count how many times the null hypothesis is rejected at the 0.01 level,
\begin{code}
> mean(p <= .01)
[1] 0.739
\end{code}
\noindent which is, unsurprisingly, quite close to the \verb|0.749| value obtained through \verb|pwr.seg()|. 

\paragraph{Sample size computation}

From the practitioner's viewpoint, it is probably more useful to compute the appropriate sample size corresponding to the specified power level. When the argument \texttt{pow} is filled in, the function returns the (rounded) sample size value. For instance, assuming a segmented variable having a Normal distribution $\mathcal{N}(5,1.5)$, the sample size corresponding to the desired power of $0.85$ is obtained simply via  

\begin{code}
> pwr.seg(pow = .85, z = "qnorm(p, 5, 1.5)", 
>                   psi = 5.5, d = .04, s = .05)
Est. sample size: 114 
\end{code}

\paragraph{Post-experimental power computation}

The function \verb|pwr.seg()| can also be used to compute the power corresponding to a fitted segmented model.

At this aim, we use the above simulated \verb|z| and \verb|y| values.

\begin{code}
> o <- lm(y ~ z)
> os <- segmented(o)
> pwr.seg(os)
Est. power for the current fit: 0.867 
\end{code}

Confidence interval replicates can be drawn to build a 95\% `confidence interval' for the power through the \texttt{ci.pow} argument.

\begin{code}
> pwr.seg(os, ci.pow = 500)
Est. power for the current fit: 0.867  ( 0.158, 0.998 ) 
\end{code}
The endpoints ($0.158$ and $0.998$) have been obtained as the $95\%$ quantiles of the power values obtained by using 500 values of slope difference and changepoint generated from a Bivariate Normal distribution with mean $(\hat{\delta},\hat{\psi})$ and corresponding covariance matrix $\mathrm{Var}(\hat{\delta},\hat{\psi})$.

\subsection{Shiny app 
}\label{sec:shiny}

The above-illustrated code is implemented with a more user-friendly interface in a Shiny app available at \url{https://uy1z3u-nicoletta0d0angelo.shinyapps.io/power_seg/}.

\begin{figure}[htb!]
	\centering
	\includegraphics[width=\textwidth]{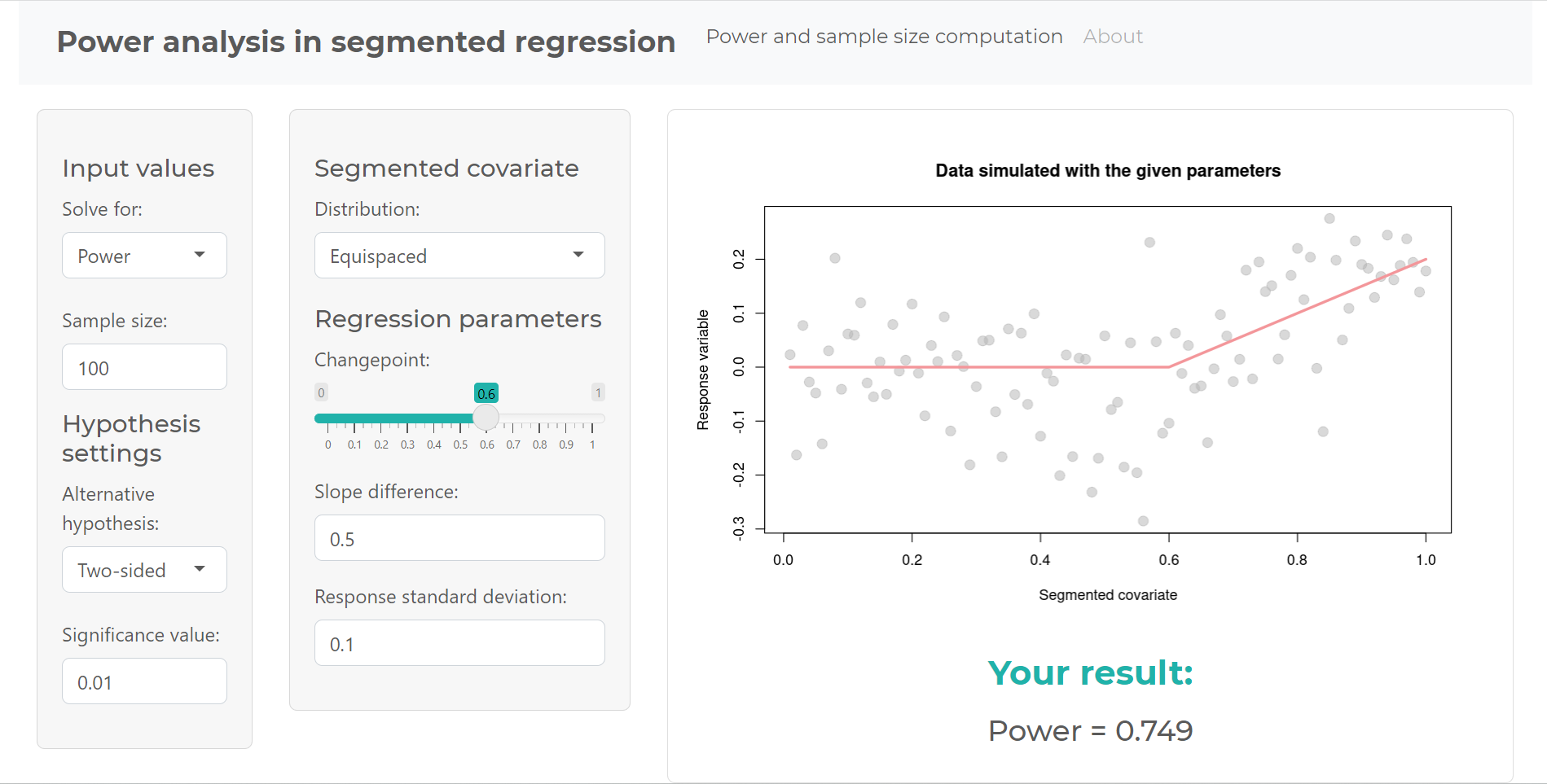}
	\caption{Shiny app interface.}
	\label{fig:shiny}
\end{figure}

The shiny app represents a tool for carrying out power analysis and sample size calculation in the context of segmented regression models, allowing one to compute the power corresponding to the specified sample size or to compute the sample size corresponding to the specified power. Figure \ref{fig:shiny} shows the interface.

First, the user should specify whether they want to obtain a sample size given a desired power value or the estimated power given a fixed sample size.
Depending on the first choice, the desired power or sample size should be imputed.

The “Alternative hypothesis” box allows you to choose among the alternatives “Two-sided” (the default), “Greater”, and “Less”.

Moreover, the significance value, set to a default of 0.01, can be changed.

The distribution of the covariate understood to have a segmented effect can be chosen among “Normal”, “Uniform”, and “Exponential”. The default is to “Equispaced” values in 0 and 1.
The parameters of the selected distribution can be changed, but the changepoint value should be modified accordingly.

Finally, also the slope difference and the response standard deviation should be imputed as desired (default to 0.5 for the normal distribution, and 0.1 otherwise).

As a result, power or sample size is displayed, and a simulated dataset (using the set parameters) is portrayed.

\section{Conclusions}
\label{sec:conc}

The growing interest in psychometric problems which can be solved by TFCPs continues to motivate researchers to develop methods for changepoint detection. 


This paper proposes an approach which relies on the pseudo Score statistics. 
The novel method is based on the pseudo Score statistics previously introduced in \cite{muggeo2016testing}, for which the sampling distribution under the alternative hypothesis has been determined. 


Through a simulation study, we proved that such test statistic achieves better performance if compared to the standard TFCPs methods for both Gaussian distributed and binary data, covering a wide range of possible applications in psychometry.

Moreover, the proposed method benefits from further advantages: being simple to compute, without resampling or simulations requested to get critical values, and with a known null/alternative distribution, making straightforward computation of the power level and/or sample size planning. 

The knowledge of the alternative distribution of our proposed statistics allows us to deal with power analysis, which is crucial when the researcher needs to estimate either the sample size or the power level of a study where the data are assumed to exhibit a piecewise relationship with an unknown changepoint.
Given those results, we advocate the use of the pseudo Score statistics to perform power analysis.

In particular, we implemented the method in an \texttt{R} function contained in the \texttt{segmented} package \citep{muggeo2008segmented}, as well as in a Shin yApp, covering the different scenarios, for instance, implementing the most common distributions that could be assumed for the data.

All in all, the proposal supports the well-established idea that statistical power calculations can be valuable in planning an experiment. Nevertheless, as shown in the illustration of the codes, we also implemented the possibility of running post-experiment power calculations, should this be used to aid in interpreting the experimental results.


\bibliography{segmented}

\begin{thebibliography}{}

\bibitem[Allalouf, 2007]{allalouf2007quality}
Allalouf, A. (2007).
\newblock Quality control procedures in the scoring, equating, and reporting of
  test scores.
\newblock {\em Educational Measurement: Issues and Practice}, 26(1):36--46.

\bibitem[Andrews, 1993]{andrews1993tests}
Andrews, D.~W. (1993).
\newblock Tests for parameter instability and structural change with unknown
  change point.
\newblock {\em Econometrica: Journal of the Econometric Society}, pages
  821--856.

\bibitem[Betts et~al., 2007]{betts2007thresholds}
Betts, M.~G., Forbes, G.~J., and Diamond, A.~W. (2007).
\newblock Thresholds in songbird occurrence in relation to landscape structure.
\newblock {\em Conservation Biology}, 21(4):1046--1058.

\bibitem[Chen and Gupta, 2012]{chen2012parametric}
Chen, J. and Gupta, A.~K. (2012).
\newblock {\em Parametric statistical change point analysis: with applications
  to genetics, medicine, and finance}.
\newblock Springer.

\bibitem[Cheng and Shao, 2021]{cheng2021application}
Cheng, Y. and Shao, C. (2021).
\newblock Application of change point analysis of response time data to detect
  test speededness.
\newblock {\em Educational and Psychological Measurement}, doi:
  10.1177/00131644211046392.

\bibitem[Cohen, 2013]{cohen2013statistical}
Cohen, J. (2013).
\newblock {\em Statistical power analysis for the behavioral sciences}.
\newblock Academic press.

\bibitem[Cs{\"o}rg{\H{o}} et~al., 1997]{csorgHo1997integral}
Cs{\"o}rg{\H{o}}, M., Horv{\'a}th, L., and Szyszkowicz, B. (1997).
\newblock Integral tests for suprema of kiefer processes with application.
\newblock {\em Statistics \& Risk Modeling}, 15(4):365--378.

\bibitem[Green and MacLeod, 2016]{green2016simr}
Green, P. and MacLeod, C.~J. (2016).
\newblock Simr: an r package for power analysis of generalized linear mixed
  models by simulation.
\newblock {\em Methods in Ecology and Evolution}, 7(4):493--498.

\bibitem[Hawkins et~al., 2003]{hawkins2003changepoint}
Hawkins, D.~M., Qiu, P., and Kang, C.~W. (2003).
\newblock The changepoint model for statistical process control.
\newblock {\em Journal of quality technology}, 35(4):355--366.

\bibitem[Kooperberg and Hsu, 2015]{powerGWASinteraction}
Kooperberg, C. and Hsu, L. (2015).
\newblock {\em powerGWASinteraction: Power Calculations for GxE and GxG
  Interactions for GWAS}.
\newblock R package version 1.1.3.

\bibitem[Lee and Jia, 2014]{lee2014using}
Lee, Y.-H. and Jia, Y. (2014).
\newblock Using response time to investigate students' test-taking behaviors in
  a naep computer-based study.
\newblock {\em Large-scale Assessments in Education}, 2(1):1--24.

\bibitem[Lee and von Davier, 2013]{lee2013monitoring}
Lee, Y.-H. and von Davier, A.~A. (2013).
\newblock Monitoring scale scores over time via quality control charts,
  model-based approaches, and time series techniques.
\newblock {\em Psychometrika}, 78(3):557--575.

\bibitem[Moerbeek, 2022]{Moerbeek2022}
Moerbeek, M. (2022).
\newblock Power analysis of longitudinal studies with piecewise linear growth
  and attrition.
\newblock {\em Behavior Research Methods}.

\bibitem[Montgomery, 2020]{montgomery2020introduction}
Montgomery, D.~C. (2020).
\newblock {\em Introduction to statistical quality control}.
\newblock John Wiley \& Sons.

\bibitem[Muggeo, 2003]{muggeo2003estimating}
Muggeo, V.~M. (2003).
\newblock Estimating regression models with unknown break-points.
\newblock {\em Statistics in Medicine}, 22(19):3055--3071.

\bibitem[Muggeo, 2016]{muggeo2016testing}
Muggeo, V.~M. (2016).
\newblock Testing with a nuisance parameter present only under the alternative:
  a score-based approach with application to segmented modelling.
\newblock {\em Journal of Statistical Computation and Simulation},
  86(15):3059--3067.

\bibitem[Muggeo, 2008]{muggeo2008segmented}
Muggeo, V. M.~R. (2008).
\newblock segmented: An {R} package to fit regression models with broken-line
  relationships.
\newblock {\em R News}, 8/1:20--25.

\bibitem[Priulla et~al., 2021]{priulla2021analysis}
Priulla, A., D’Angelo, N., and Attanasio, M. (2021).
\newblock An analysis of italian university students’ performance through
  segmented regression models: gender differences in stem courses.
\newblock {\em Genus}, 77:1--20.

\bibitem[Qiu et~al., 2021]{powerSurvEpi}
Qiu, W., Chavarro, J., Lazarus, R., Rosner, B., and Ma., J. (2021).
\newblock {\em powerSurvEpi: Power and Sample Size Calculation for Survival
  Analysis of Epidemiological Studies}.
\newblock R package version 0.1.3.

\bibitem[{R Core Team}, 2024]{R}
{R Core Team} (2024).
\newblock {\em R: A Language and Environment for Statistical Computing}.
\newblock R Foundation for Statistical Computing, Vienna, Austria.

\bibitem[Rasch, 1993]{rasch1993probabilistic}
Rasch, G. (1993).
\newblock {\em Probabilistic models for some intelligence and attainment
  tests.}
\newblock ERIC.

\bibitem[Shao et~al., 2015]{shao2015change}
Shao, C., Li, J., and Cheng, Y. (2015).
\newblock A change point based method for test speededness detection.
\newblock {\em Psychometrika. doi}, 10:1007.

\bibitem[Sinharay, 2016]{sinharay16}
Sinharay, S. (2016).
\newblock Person fit analysis in computerized adaptive testing using tests for
  a change point.
\newblock {\em Journal of Educational and Behavioral Statistics}, 41:521--549.

\bibitem[Sinharay, 2017]{sinharay17}
Sinharay, S. (2017).
\newblock Some remarks on applications of tests for detecting a change point to
  psychometric problems.
\newblock {\em Psychometrika}, 82:1149--1161.

\bibitem[Ulm, 1991]{ulm1991statistical}
Ulm, K. (1991).
\newblock A statistical method for assessing a threshold in epidemiological
  studies.
\newblock {\em Statistics in Medicine}, 10(3):341--349.

\bibitem[van~der Linden and Xiong, 2013]{van2013speededness}
van~der Linden, W.~J. and Xiong, X. (2013).
\newblock Speededness and adaptive testing.
\newblock {\em Journal of Educational and Behavioral Statistics},
  38(4):418--438.

\bibitem[von Davier, 2012]{von2012use}
von Davier, A.~A. (2012).
\newblock The use of quality control and data mining techniques for monitoring
  scaled scores: An overview.
\newblock {\em ETS Research Report Series}, 2012(2):i--18.

\bibitem[Worsley, 1979]{worsley1979likelihood}
Worsley, K. (1979).
\newblock On the likelihood ratio test for a shift in location of normal
  populations.
\newblock {\em Journal of the American Statistical Association},
  74(366a):365--367.

\bibitem[Zhu et~al., 2023]{zhu2023bayesian}
Zhu, H., Jiao, H., Gao, W., and Meng, X. (2023).
\newblock Bayesian change-point analysis approach to detecting aberrant
  test-taking behavior using response times.
\newblock {\em Journal of Educational and Behavioral Statistics},
  48(4):490--520.

\end{thebibliography}

\end{document}